\documentclass[11pt]{article}
\usepackage{Blois,epsfig}
\usepackage{amssymb}
\usepackage{amsmath}

\def\be{\begin{equation}}
\def\ee{\end{equation}}
\def\bea{\begin{eqnarray}}
\def\eea{\end{eqnarray}}

\newcommand{\g}{\gamma}
\newcommand{\f}{\frac}

\newcommand{\pom}{{\mathbb P}}

\newcommand{\xp}{x_{\mathbb P}}

\newcommand\lr[1]{{\left({#1}\right)}}
\begin{document}
\vspace*{2cm}

\vspace*{2cm}
\title{TESTING SATURATION WITH LARGE-MASS DIFFRACTION IN DIS}

\author{C. MARQUET}
\address{Service de Physique Th{\'e}orique, CEA/Saclay,
91191 Gif-sur-Yvette Cedex, France\\
URA 2306, unit{\'e} de recherche associ{\'e}e au CNRS}

\author{K. GOLEC-BIERNAT}
\address{Institute of Nuclear Physics, Radzikowskiego 152,
31-342 Krak{\'o}w, Poland\\
Institute of Physics, University of Rzesz\'ow, Rzesz\'ow, Poland}

\maketitle
\abstracts{In large-mass diffraction in deep inelastic scattering, we study the 
gluon jet close to the rapidity gap and show that its transverse momentum 
spectrum is peaked around a value which caracterizes the onset of S-matrix 
unitarity. We argue that such a measurement of diffractive jets could help 
understanding if at the present energies, unitarity comes as a consequence of 
saturation.}

\section{Introduction}

One of the most successful approaches to the understanding of hard diffraction 
in DIS is the QCD dipole picture which naturally describes both inclusive and 
diffractive events within the same theoretical 
framework~\cite{nikzak,biapesroy}. It expresses the scattering of the photon 
through its fluctuation into a color singlet $q\bar q$ pair (dipole) and the 
same dipole scattering amplitude enters in the formulation of the inclusive and 
diffractive proton structure functions. This approach revealed~\cite{golec} that 
unitarity and the way it is realized should be important ingredients of the 
description of diffractive cross-sections, making those ideal places to look for 
saturation effects at small-$x$.

In this study, we analyse hard diffraction when the proton stays intact after 
the collision and the mass of the diffractive final state is much bigger
than the virtuality of the photon. This process is called diffractive photon 
dissociation. We propose the measurement of the final state configuration ${\rm 
X+jet+gap+p}$ in virtual photon-proton collisions. We show that in the context 
of saturation theory, the transverse momentum distribution of the measured jet 
is resonant with the scale at which the contributions of large-size dipoles 
start to be suppressed, called the saturation scale. That exhibits the potential 
of the diffractive-jet measurement for testing saturation and extracting the 
saturation scale.

\section{Diffractive photon dissociation}

\begin{figure}[htb]
\hspace{0.5cm}
\begin{minipage}[t]{55mm}
\includegraphics[width=5cm]{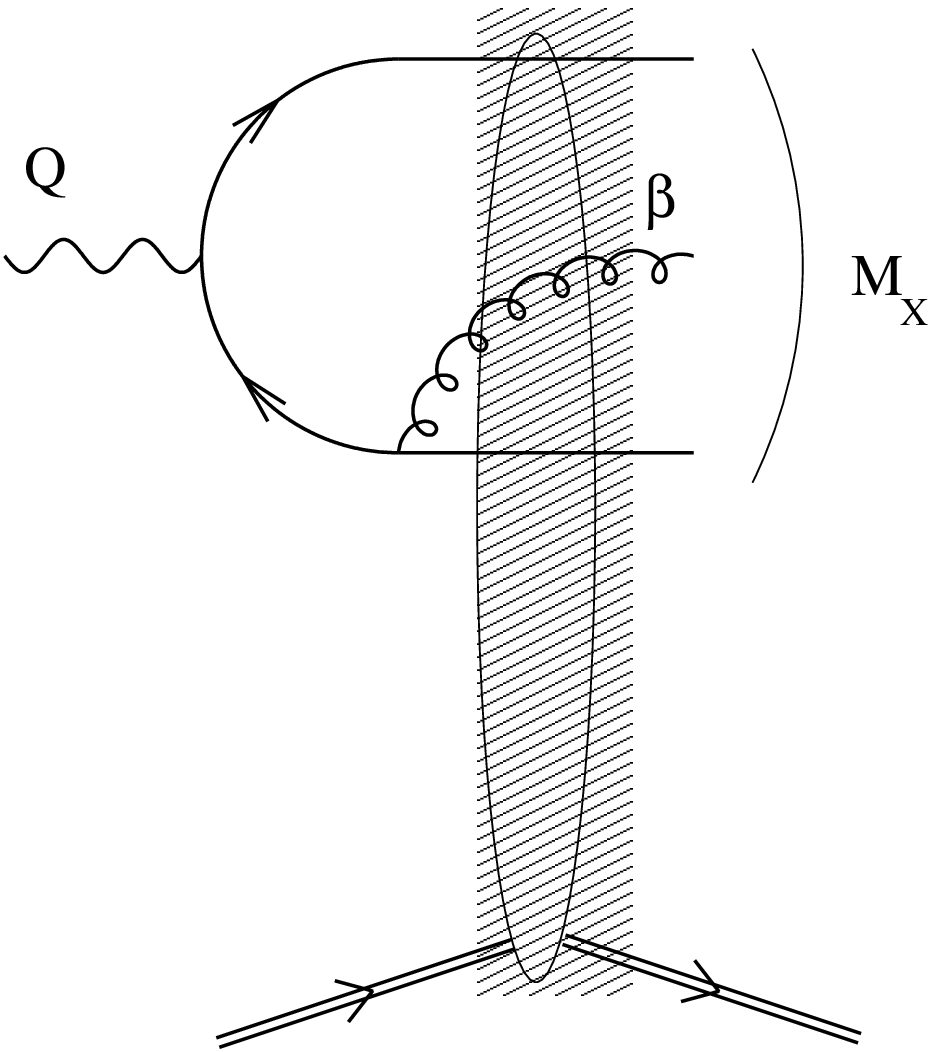}
\end{minipage}
\hspace{\fill}
\begin{minipage}[t]{75mm}
\includegraphics[width=6.8cm]{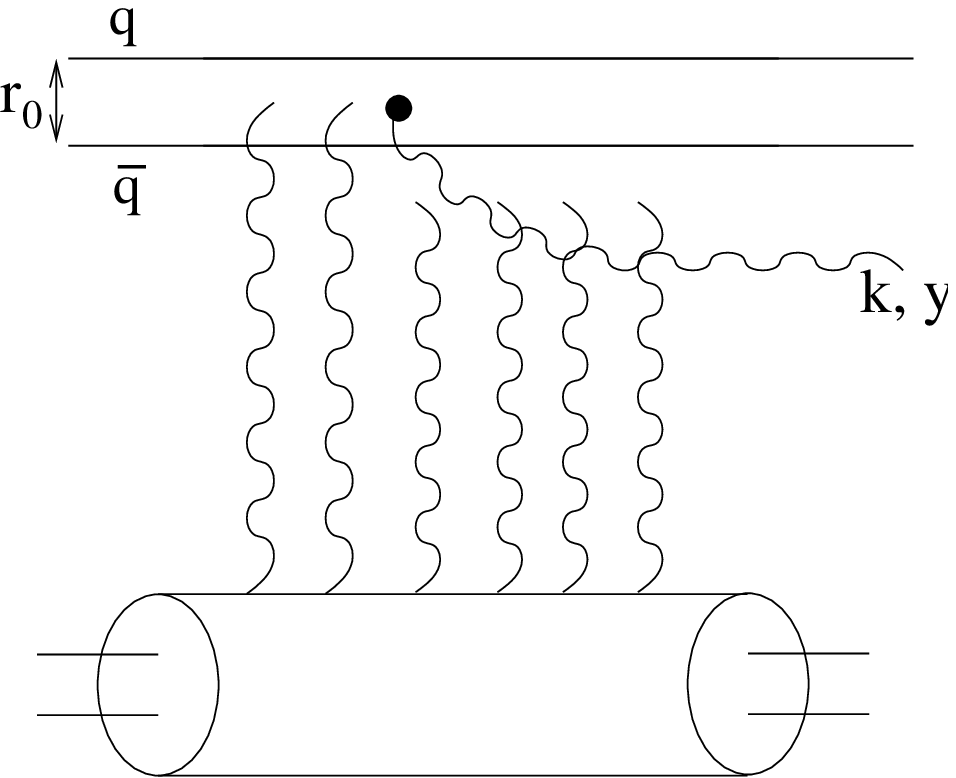}
\end{minipage}
\hspace{0.5cm}
\caption{Left side: diffractive dissociation of a photon of virtuality $Q$ into 
a final state of mass $M_X.$ Right side: diffractive production of a gluon with 
transverse momentum $k$ and rapidity $y\!=\!\log{1/\beta}$ off a $q\bar q$ 
dipole of size $r_0.$}
\label{F1}
\end{figure}

In deep inelastic scattering, a photon of virtuality $Q^2$ collides
with a proton. In an appropriate frame called the dipole frame, the virtual 
photon undergoes the hadronic interaction via a fluctuation into a dipole; the 
dipole then interacts with the target proton and one has the following 
factorization
\be
\sigma^{\g^*p}=\int d^2 r 
\int_0^1 d\alpha
\left(|\psi_T^{\g}(r,\alpha;Q)|^2
+|\psi_L^{\g}(r,\alpha;Q)|^2
\right) \sigma(r)\label{factor}
\ee
which relates a cross-section for an incident photon $\sigma^{\g^*p}$ to the 
corresponding cross-section $\sigma(r)$ for an incident dipole of transverse 
size $r.$ In the leading logarithmic approximation we are interested in, the 
dipole cross-sections do not depend on $\alpha,$ the fraction of photon 
longitudinal momentum carried by the antiquark. The wavefunctions 
$|\psi^\gamma_T|^2$ and 
$|\psi^\gamma_L|^2$ describing the splitting of the transversely (T) or 
longitudinally (L) polarized photon on the dipole are well known. 

In diffractive deep inelastic scattering, the proton gets out of the collision 
intact and there is a rapidity gap between that proton and the final state $X,$ 
see the left plot of Fig.~1. If the final state diffractive mass $M_X$ is much 
bigger than 
$Q,$ then the dominant configurations of the final state come from the $q\bar q 
g$ component of the photon wavefunction or from higher Fock states, 
i.e. from the photon dissociation. In this study, we consider this kinematical 
regime where $\beta\!\equiv\!Q^2/(Q^2+M_X^2)\!\ll\!1$ and we investigate 
the $q\bar q g$ component. In this case, due to the infrared singularity of QCD, 
the final-state configuration that gives the dominant contribution to the 
cross-section corresponds to the gluon jet being the closest to the gap. We 
shall study the transverse momentum spectrum of this gluon jet within 
high-energy QCD.

The right plot of Fig.~1 represents the diffractive production of a gluon with 
transverse momentum $k$ and rapidity $y$ in the collision of a dipole of 
tranverse size $r_{0}$ with the target proton. Provided $k$ is a perturbative 
scale, the corresponding cross-section reads~\cite{cyrille}
\be
\label{eq:sigd}
\f{d\sigma^{diff}}{d^2kdy}(r_{0})\,=\,\f{\alpha_sN_c^2}{4\pi^2C_F}
\int d^2b\ A(k,x_0,x_1;\Delta\eta)\cdot A^*(k,x_0,x_1;\Delta\eta)
\ee
where $x_0\!=\!b\!+\!r_{0}/2$, $x_1\!=\!b\!-\!r_{0}/2$, $Y$ is the total 
rapidity and $\Delta\eta\!=\!Y\!-\!y$ is the rapidity gap. The two-dimensional 
vector $A$ is given by
\be
A(k,x_0,x_1;\Delta\eta)=\int\f{d^2z}{2\pi}\ e^{-ik\cdot z}
\left[\f{z\!-\!x_0}{|z\!-\!x_0|^2}-\f{z\!-\!x_1}{|z\!-\!x_1|^2}\right]
\lr{S^{(2)}(x_0,z,x_1;\Delta\eta)-S(x_0,x_1;\Delta\eta)}
\label{ampla}
\ee
where $S(x_0,x_1;\Delta\eta)$ is the elastic ${\cal S}-$matrix for the collision 
of the 
dipole $(x_0,x_1)$ on the target proton evolved at the rapidity $\Delta\eta$, 
and 
$S^{(2)}(x_0,z,x_1;\Delta\eta)$ is the elastic ${\cal S}-$matrix for the 
collision of two dipoles $(x_0,z)$ and $(z,x_1)$. Those QCD ${\cal S}-$matrices
contain both perturbative and non-perturbative information.


We are going to study the observable
\be
k^2 \f{d\sigma_{diff}^\g}{d^2k\ dM_X}
\label{obs}
\ee
which is differential with respect to the diffractive mass $M_X$ and to the 
final-state gluon transverse momentum $k.$ The cross-section 
(\ref{obs}) is obtained from formulae (\ref{factor}) and (\ref{eq:sigd}), along 
with the usual kinematics of diffractive DIS: $Y\!=\!\log(1/x)$ and
$\Delta\eta\!=\!\log(1/x_{\pom})$ with $x\!=\!Q^2/(Q^2+W^2)$ and 
$x_{\pom}\!=\!x/\beta.$ $W^2$ is the center-of-mass energy of the photon-proton 
collision.
Independently of the form of the ${\cal S}-$matrices, one can show~\cite{us} 
that the behavior of the observable (\ref{obs}) 
as a function of the gluon transverse momentum $k$ is the following:
it is going to rise as $k^2$ for small values of $k$ and
fall as $1/k^2$ for large values of $k$. A maximum will occur for a value
$k_0$ which is related to the inverse of the typical size for which the ${\cal
S}-$matrices approach zero; in other words, the maximum $k_0$ will reflect the 
scale at which unitarity sets in. The question is whether unitarity is due to 
non-perturbative physics or if it rather comes as a consequence of parton 
saturation. We shall explore the latter possibility, in which case $k_0$ is 
identified to the saturation scale $Q_s.$

\section{Phenomenology}

\begin{figure}[t]
\hspace{0.5cm}
\begin{minipage}[t]{65mm}
\includegraphics[width=6.2cm]{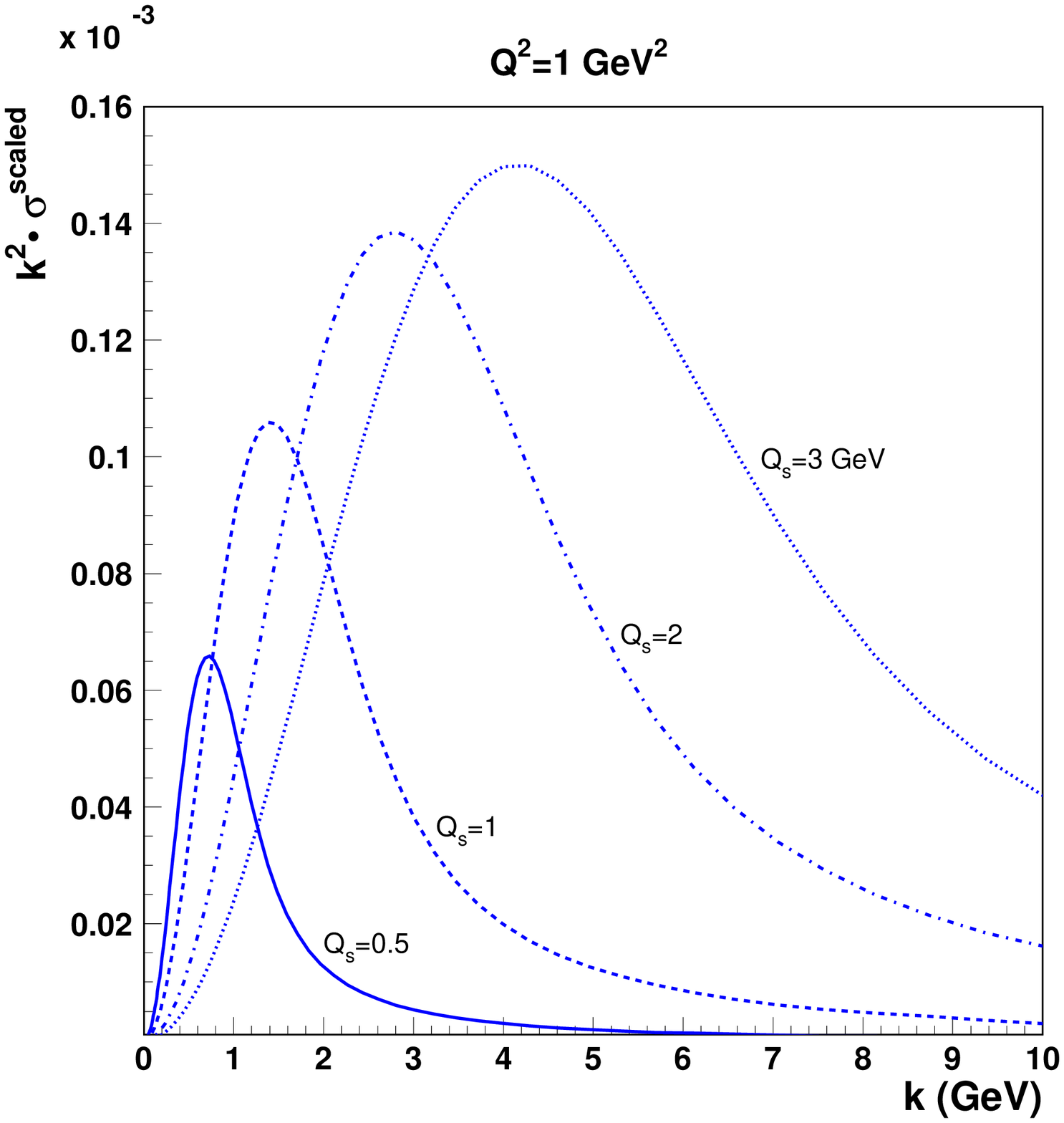}
\label{gluprod}
\end{minipage}
\hspace{\fill}
\begin{minipage}[t]{70mm}
\includegraphics[width=6.9cm]{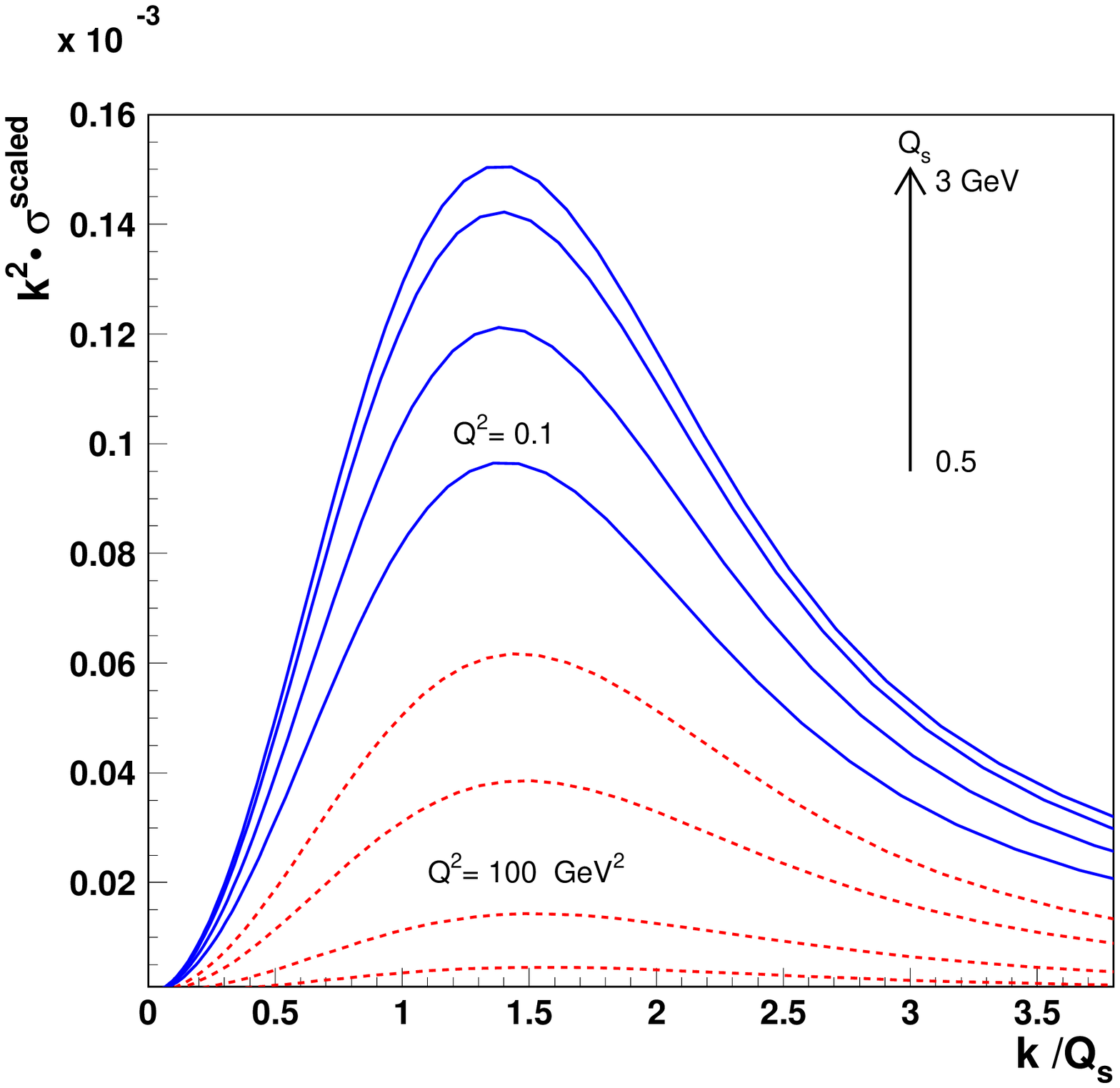}
\label{F2}
\end{minipage}
\hspace{0.5cm}
\caption{The observable (\ref{obs}) as a function of the gluon transverse 
momentum $k.$ Left plot: for a fixed value of $Q^2\!=\!1~{\rm GeV}^2$ and four 
indicated values of the saturation scale $Q_s$. Right plot: as a function of 
$k/Q_s$ for two extreme values of $Q^2$ equal to $0.1$ and $100~{\rm GeV}^2$ and 
the four values of the saturation scale used for the left plot.}
\end{figure}

The exact form of the ${\cal S}-$matrices is unknown, and we have to consider 
models in order to produce values of the observable (\ref{obs}) at any value of 
$k$. For this purpose we consider the following models, inspired by the GBW 
parametrization \cite{golec} of parton saturation effects:
\bea
\label{eq:S1}
S(x_0,x_1;\Delta\eta)\!&=&\!\Theta(R_p\!-\!|b|)\,\,{\rm 
e}^{-Q_s^2(x_{\pom})(x_0\!-\!x_1)^2/4}
\,+\,\Theta(|b|\!-\!R_p)\ ,
\\
\label{eq:S2}
S^{(2)}(x_0,z,x_1;\Delta\eta)\!&=&\!\Theta(R_p\!-\!|b|)\,\,
{\rm e}^{-Q_s^2(x_{\pom})(x_0\!-\!z)^2/4}\,\,
{\rm e}^{-Q_s^2(x_{\pom})(z\!-\!x_1)^2/4}\,+\,\Theta(|b|\!-\!R_p)\ ,
\eea
where $R_p$ is the radius of the proton. The saturation scale $Q_s$, the 
basic quantity characterizing saturation effects, is a rising function of 
energy through its $\xp-$dependence; this can be tested using the observable
(\ref{obs}) and the phenomenon explained above.

In Fig.~2 we plot the observable (\ref{obs}) as a function of $k.$ The left plot 
is with fixed $Q^2\!=\!1~{\rm GeV}^2$ and four values of the saturation scale,
$Q_s\!=\!0.5, 1, 2, 3~{\rm GeV}$. As discussed in section 2, we check that 
$k^2 d\sigma_{diff}^\g/d^2k dM_X$ grows as $k^2$ at small momentum and decrases 
as $1/k^2$ at large momentum. We also see that the transition region between the 
two distinct behaviors at small and large $k$, which features a marked bump, is 
linked to the value of $Q_s$ as expected. This is confirmed on the right plot 
where the same observable is plotted as a function of the rescaled transverse 
momentum $k/Q_s$ for two extreme values of the photon virtuality, $Q^2\!=\!0.1$ 
and $100~{\rm GeV}^2$. Clearly the maximum for each curve is independent of 
$Q_s$ and $Q^2$ in a broad range of considered values: one has 
$k_{max}/Q_s\!\sim\!1.4.$ Therefore 
diffractive gluon production in the domain of large diffractive mass offers a 
unique opportunity to determine the saturation scale $Q_s$ and its dependence on 
$\xp.$ Note that since 
$k_{max}$ is independent of $Q^2,$ a wide range of photon virtuality could be 
used to carry out this measurement, as long as one keeps $\beta\!\ll\!1.$

\begin{figure}[t]
\begin{center}
\epsfig{file=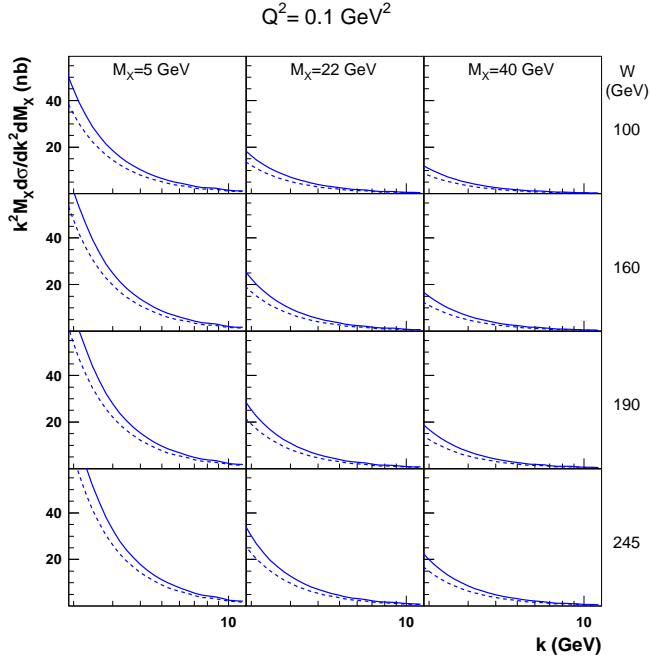,width=8.5cm}
\caption{The cross-section $k^2M_Xd\sigma/dk^2dM_X$ as a function of the jet 
transverse momentum $k$ for $Q^2\!=\!0.1~{\rm GeV}^2$ and different values of 
diffractive mass $M_X$ and energy $W$. Full lines: light quarks only, dashed 
lines: with charm.}
\end{center}
\label{F5}
\end{figure}

However, from the experimental point of view there exists an important 
limitation related to the minimal value of the transverse momentum which can 
be measured for a jet. In the most pesymistic scenario, considering even rather 
high values of the saturation scale, $Q_s(\xp)\sim\!1~{\rm GeV}$, it is unlikely
that the maximum $k_{max}$ of the cross section (\ref{obs}) can be seen at HERA.
Thus, to see the transition between the two different behaviors of the cross
section (\ref{obs}) seems like a major experimental challenge. In Fig.~3 we 
illustrated such a situation, where the cross-section $k^2M_Xd\sigma/dk^2dM_X$ 
is plotted in the HERA energy range with the saturation scale taken from 
Ref.~\cite{golec} and the overall normalization $\alpha_s\!=\!0.15$ taken from 
Ref.~\cite{munsho}. One sees that, unfortunately, the data should always lie on 
the perturbative side of the bump. However, it is not necessary to see the whole 
bump to confirm the influence of the saturation scale on the results. In 
particular, there is a big difference in the rise towards the bump between the 
highest $\xp-$bin ($M_X\!=\!40~{\rm GeV}$ and $W\!=\!100~{\rm GeV}$) and the 
lowest $\xp-$bin ($M_X\!=\!5~{\rm GeV}$ and $W\!=\!245~{\rm GeV}$). An 
experimental confirmation of such a behavior would be in favor of saturation and 
could lead to the determination of the saturation scale. If however this trend 
is not observed, it could mean that in this process, unitarity does not come 
from saturation, but rather from soft physics.

\end{document}